\documentclass[aps,preprint,superscriptaddress]{revtex4-1}  % showing figures

\pdfoutput=1
\usepackage{xr}

\usepackage{graphicx}
\usepackage[space]{grffile}
\usepackage{latexsym,color}
\usepackage{textcomp}
\usepackage{longtable}
\usepackage{multirow,booktabs}
\usepackage{amsfonts,amsmath,amssymb}
\usepackage{siunitx}
\usepackage[version=4]{mhchem}
\usepackage{natbib}
\usepackage{url}
\usepackage{hyperref}
\hypersetup{colorlinks=false,pdfborder={0 0 0}}
\newif\iflatexml\latexmlfalse
\usepackage[utf8]{inputenc}
\usepackage[english]{babel}

\graphicspath{ {./figures/} }

\begin{document}

\title{Dynamics-dependent density distribution in active suspensions}

%\author{Jochen Arlt$^{1, *}$, Vincent A Martinez$^1$, Angela Dawson$^1$, Teuta Pilizota$^2$, \& Wilson C K Poon$^1$}
\author{Jochen Arlt}
\email{j.arlt@ed.ac.uk}
\thanks{Corresponding author}
\affiliation{School of Physics \& Astronomy, The University of Edinburgh, Peter Guthrie Tait Road, Edinburgh EH9 3FD, United Kingdom}

\author{Vincent A Martinez}
\affiliation{School of Physics \& Astronomy, The University of Edinburgh, Peter Guthrie Tait Road, Edinburgh EH9 3FD, United Kingdom}

\author{Angela Dawson}
\affiliation{School of Physics \& Astronomy, The University of Edinburgh, Peter Guthrie Tait Road, Edinburgh EH9 3FD, United Kingdom}

\author{Teuta Pilizota}
\affiliation{School of Biological Sciences and Centre for Synthetic and Systems Biology, The University of Edinburgh, Alexander Crum Brown Road, Edinburgh EH9 3FF, United Kingdom}

\author{Wilson C K Poon}
\affiliation{School of Physics \& Astronomy, The University of Edinburgh, Peter Guthrie Tait Road, Edinburgh EH9 3FD, United Kingdom}

\begin{abstract}
Self-propelled colloids constitute an important class of intrinsically non-equilibrium matter. Typically, such a particle moves ballistically at short times, but eventually changes its orientation, and displays random-walk behavior in the long-time limit. Theory predicts that if the velocity of non-interacting swimmers varies spatially in 1D, $v(x)$, then their density $\rho(x)$ satisfies $\rho(x) = \rho(0)v(0)/v(x)$, where $x = 0$ is an arbitrary reference point. Such a dependence of steady-state $\rho(x)$ on the particle dynamics, which was the qualitative basis of recent work demonstrating how to `paint' with bacteria, is forbidden in thermal equilibrium. We verify this prediction quantitatively by constructing bacteria that swim with an intensity-dependent speed when illuminated. A spatial light pattern therefore creates a speed profile, along which we find that, indeed, $\rho(x)v(x) = $~constant, provided that steady state is reached.  
\end{abstract}

% Keywords are not mandatory, but authors are strongly encouraged to provide them. If provided, please include two to five keywords, separated by the pipe symbol, e.g:

\maketitle

Einstein predicted and Perrin verified that, in a gravitational field, the equilibrium (number) density of a dilute dispersion of colloidal particles varied with height $z$ according to 
\begin{equation}
\rho (z) = \rho(0) e^{-z/z_0}, \label{eq:Perrin}
\end{equation}
where $z_0$ encodes the equality of diffusive and sedimentation fluxes:
\begin{equation}
z_0 = \frac{D_0}{v_{\rm s}}, \label{eq:height}
\end{equation}
with $v_{\rm s}$ a particle's sedimentation speed and $D_0$ its thermal diffusivity. For spheres of radius $a$ in a liquid of viscosity $\eta_0$ at temperature $T$ with density $\Delta$ lower that of the particles $v_{\rm s} = 2ga^2\Delta/9\eta_0$ and
$D_0 = k_{\rm B}T/6\pi\eta_0 a$ where $k_{\rm B}$ is Boltzmann's constant, so that $z_0 = 3k_{\rm B}T/4\pi ga^3 \Delta$. The verification of Eq.~\ref{eq:Perrin} demonstrated the granularity of matter \cite{PerrinAtoms}. 

Now suppose $\eta_0 = \eta_0(z)$, so that the particle dynamics also varies in space: $D_0 = D_0(z)$. Nevertheless, statistical mechanics stipulates that  the spatially-dependent dynamical coefficient $D_0(z)$ {\it cannot} appear in the {\it equilibrium} density distribution, and Eq.~\ref{eq:Perrin} still holds because the $\eta$-dependence cancels out in Eq.~\ref{eq:height}. 
 
Active colloids \citep{poon2013physics}, particles that dissipate energy to propel themselves, form an important class of active matter \citep{Sriram,Joanny2014}. Such dissipative states of matter, which include all living organisms, are intrinsically non-equilibrium, and give rise to new physics. Consider a system of run-and-tumble particles (RTPs). An RTP self propels (`runs') at velocity $\mathbf{v}$ for time $\tau_{\rm run}$, then changes direction (`tumbles') instantaneously to run at $\mathbf{v}^{\prime}$ such that $|\mathbf{v}^\prime| = |\mathbf{v}|$ but with randomised direction, so that at times $\gg \tau_{\rm run}$ it behaves as a random walker. Suppose the run speed of such particles is spatially dependent, $v(\mathbf{r})$. Solving a Fokker-Planck equation for the coarse-grained kinetics of RTPs, Tailleur and Cates \citep{TailleurPRL,TailleurEPL} predicted that the resulting density distribution should be $\rho(\mathbf{r}) = \rho(\mathbf{0})v(\mathbf{0})/v(\mathbf{r})$, with $\mathbf{0}$ an arbitrary origin. (A purely mechanical derivation is also possible \citep{Brady2015}.) The appearance of the particles' dynamics, $v(\mathbf{r})$, in this formula contrasts starkly with the sedimentation equilibrium of passive colloids with spatially-dependent diffusivity $D_0(x)$, for which the $D$-independent Eq.~\ref{eq:Perrin} still holds. 

When restricted to 1D, this result becomes
\begin{equation}
\rho(x) = \rho(0)v(0)/v(x), \label{eq:rhoV}
\end{equation}
which was first derived for non-interacting random walkers by Schnitzer \citep{Schnitzer1993}. Tailleur and Cates show that it is valid for {\it interacting} RTPs whose run speed can be expressed as $v(x) = v[\rho(x)]$ \citep{TailleurPRL}. Moreover, under quite general conditions, $\rho v =$~constant also holds (provided translational diffusion is negligible) for active Brownian particles (ABPs) \citep{Cates2016}, which reorient gradually due to their rotational diffusivity $D_{\rm rot}$, losing directional memory after a persistence time of $\sim D_{\rm rot}^{-1}$. At  $t \gg D_{\rm rot}^{-1}$, an ABP is again a random walker. 

Qualitatively, Eq.~\ref{eq:rhoV} was the basis of recent demonstrations of templated self assembly using light-activated motile bacteria \cite{Jochen2018,Frangipane2018}. In a spatially-varying illumination pattern, cells accumulate in the darker regions, generating contrast. Quantitatively, however, Eq.~\ref{eq:rhoV} has remained unverified by experiments to date. 
Indeed, as its theoretical derivations do not explicitly take account of hydrodynamic interactions, it is unclear to what extent it is applicable to such systems.
Here, we quantitatively investigate eq.~\ref{eq:rhoV} with the same light-activated {\it Escherichia coli} bacteria used previously to demonstrate templated active self assembly \cite{Jochen2018}.

Each {\it E. coli} is an $\approx \SI{2}{\micro\meter} \times \SI{1}{\micro\meter}$ spherocylinder with $\approx 7$--$\SI{10}{\micro\meter}$ helical flagella powered by rotary motors \citep{BergBook}. When all flagella rotate counterclockwise (seen from behind), they bundle and propel the cell. Every $\tau_{\rm run} \sim \SI{1}{\second}$ or so, one or more flagella reverse and unbundle, causing a change in direction: wild-type (WT) cells are RTPs \citep{BergBrown}. At a typical average speed $\bar v \gtrsim \SI{10}{\micro\meter\per\second}$, they random walk with a persistence length $l_p \sim \bar v \tau_{\rm run} \sim \SI{10}{\micro\meter}$. Deleting the {\it cheY} gene prevents tumbling; cells become ABPs with $D_{\rm rot}^{-1} \sim \SI{10}{\second}$ and $l_p \sim \bar v D_{\rm rot}^{-1} \sim \SI{100}{\micro\meter}$ \citep{Wu2006}, so that at times $\gg D_{\rm rot}^{-1}$ cells random walk with $D_{\rm eff} \sim \bar v^2 D_{\rm rot}^{-1} \sim 10^3\,\si{\square\micro\meter\per\second}$. 

Our {\it E.~coli} mutants carried a plasmid expressing proteorhodopsin (PR), which pumps protons in green light \citep{Beja2000}. Cells suspended in nutrient-free motility buffer were sealed into \SI{20}{\micro\meter} high compartments and imaged using $10\times$ phase contrast microscopy.  After some minutes, $\bar v$ dropped abruptly to zero upon oxygen exhaustion \citep{Jochen2018}. Thereafter, cells only swam when illuminated in green  \citep{Walter2007,Poon2016a,Jochen2018}, with an average speed $\bar v$ that increased with the light intensity, $\mathcal{I}$, saturating at $v_{\rm max}$ beyond some $\mathcal{I}_{0}$ \citep{Jochen2018}. These are living analogs of synthetic light-activated active colloids \citep{BechingerJanus,PalacciCrystal}.

We projected a quasi-1D stepped illumination pattern 
\begin{equation}
\mathcal{I}(x,y) = \left\{
	\begin{array}{ll}
		\mathcal{I}_+ < \mathcal{I}_0, \; x<0\\ 
		 \mathcal{I}_- < \mathcal{I}_+, \; x >0 \label{eq:Ix}
	\end{array}
\right.
\end{equation}
on a field of these cells. This generated a spatial pattern of swimming speed, $\bar v(x,y)$, and cell density, $\rho(x,y)$, which we measured using spatially-resolved differential dynamic microscopy (sDDM) \citep{CerbinoDDM1,MartinezDDM,Jochen2018}. Averaging over $y$ gives $\bar v(x)$ and $\rho(x)$, which allows a direct test of Eq.~\ref{eq:rhoV}, provided that this light pattern, Eq.~\ref{eq:Ix}, generates a corresponding {\it sharp} pattern of cell run speeds:
\begin{equation}
\bar v(x) = \left\{
	\begin{array}{ll}
		\bar v_+  \hspace{.8cm}, \;x<0\\ 
		 \bar v_- < \bar v_+, \; x >0
	\end{array}
\right. . \label{eq:vStep}
\end{equation}
This requires cells that respond rapidly to changes in $\mathcal{I}$, which was found not to be the case \citep{Jochen2018} for previous PR-expressing mutants with otherwise intact metabolism \citep{Walter2007}. Indeed, a recent attempt to verify Eq.~\ref{eq:rhoV} using PR-bearing {\it E. coli} found instead (in our notation) $\rho = (a/\bar v) + b$ with positive constants $a$ and $b$. The latter was ascribed to a long $\tau_{\rm stop}$, which led to memory effects \citep{Frangipane2018}. 

\begin{figure}[t]
\begin{center}
(a)\;\includegraphics[width=.65\columnwidth]{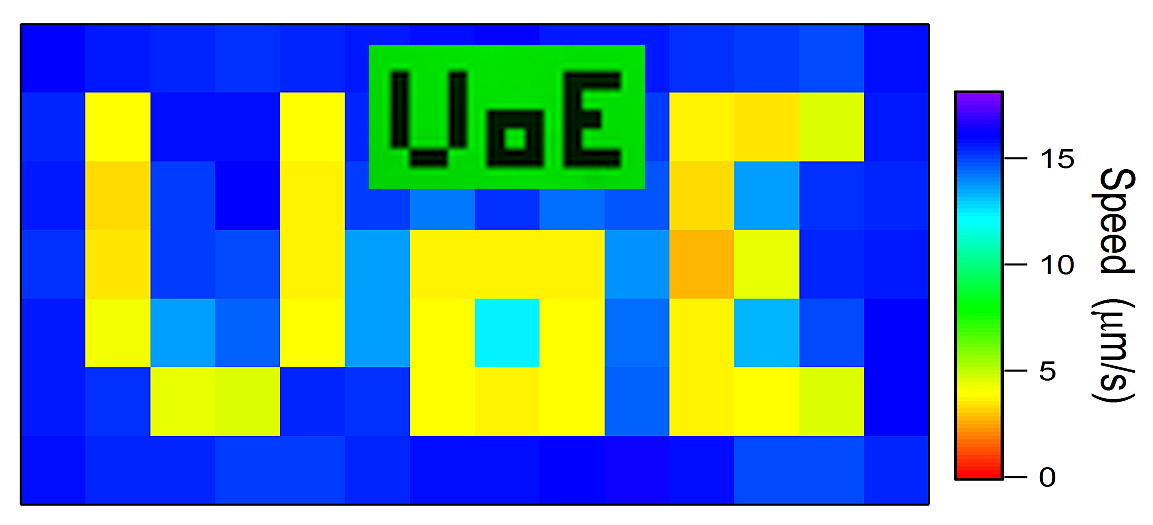}
(b)\;\includegraphics[width=.65\columnwidth]{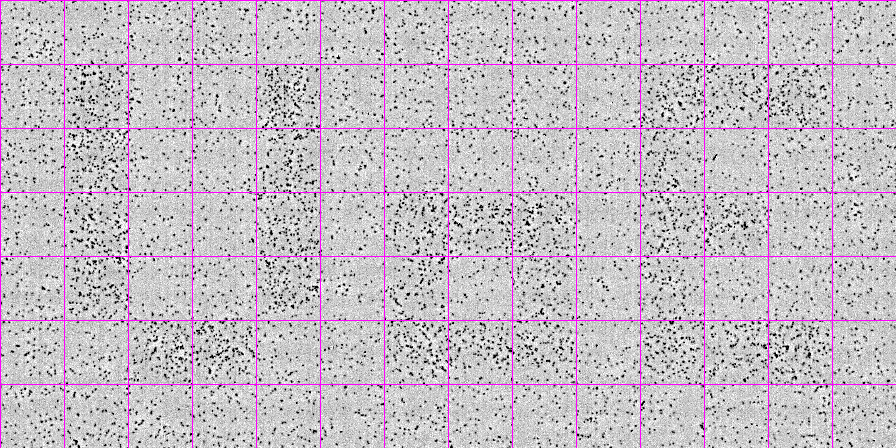}
\caption{
Projecting light patterns onto a sample of light--powered \textit{E.coli} AD10 (at OD=1) leads to a spatial variation of the mean swimming speeds $\bar{v}$. Each square tile is $90\times 90 \,\si{\square\micro\meter}$. (a) Map of $\bar{v}$ as measured by sDDM when the negative `UoE' pattern shown in inset is projected onto the sample. (b) Phase contrast image about $\SI{20}{\second}$ after applying the pattern, showing accumulation in the darker regions. Magenta lines indicate the square tiles of 64 pixels ($\SI{90}{\micro\meter}$) size used for the sDDM analysis.}
\label{fig:sDDM}
\end{center}
\end{figure}

We achieved rapid response by deleting the {\it unc} gene cluster encoding the $\textrm F_{1} \textrm F_{\textrm o}$-ATPase membrane protein complex from a parent K12-derived $\Delta${\it cheY} mutant, giving a fast-responding smooth-swimmer, AD10 \citep{Jochen2018}. 
In fully-oxygenated phosphate motility buffer (MB) at optical density OD=1, $\bar v \approx \SI{30}{\micro\meter\per\second}$ and a fraction $\beta \lesssim 20\%$ of cells were non-motile. When illuminated anaerobically, $\bar v_{\rm max} = \SI{28}{\micro\meter\per\second}$ and $\tau_{\rm stop} \ll \SI{1}{\second}$, compared to a $\tau_{\rm stop}$ of many minutes in the parent strain without {\it unc} deletion \citep{Jochen2018}. (Details of other strains we used are given in the methods section.)

%%%%%%%%%%%%%%%%%%%%%%%%%%%%%%%%%%%%%%%%%%%%%%%%%%%%%%%%%%%%%%%%%%%%
\section*{Results}

We first used a digital mirror device \citep{Jochen2018} to project a binary (bright-dark) spatial intensity pattern $\mathcal{I}(x,y)$ spelling out `UoE', Fig.~\ref{fig:sDDM}(a, inset), onto a field of cells that had been uniformly illuminating for some time, so $\bar v$ was initially constant in space. We used sDDM to measure $\bar v(x,y)$, $\beta(x,y)$ and $\rho(x,y)$ in $64\times 64$~(pixel)$^2$ tiles (see SI for details). The projected $\mathcal{I}(x,y)$ was rapidly replicated in a pattern of $\bar v(x,y)$, Fig.~\ref{fig:sDDM}(a). A similar $\rho(x,y)$ pattern soon forms, Fig.~\ref{fig:sDDM}(b), so that this effect can be used for templated self assembly \citep{Jochen2018, Frangipane2018}. Given that higher cell densities occur in darker regions with lower swimming speed, Eq.~\ref{eq:rhoV} is clearly qualitatively correct \citep{Jochen2018,Frangipane2018}. We now proceed to test it quantitatively.

Strictly speaking, a non-interacting limit does not exist for bacterial suspensions \cite{Morozov2017}. Cells interact hydrodynamically at {\it any} concentration, although simulations show that swimmers behave effectively as non-interacting when $\rho/\rho_c \lesssim 0.1$, where $\rho_c$ is the density for the onset of collective behaviour. We observed collective motion in our {\it E. coli} suspension at OD~$\gtrsim 10$, corresponding to a cell body volume fraction of $\phi\gtrsim 1.4\%$, consistent with a previous estimate of 2\% \cite{Clement2014,Morozov2017}, so that a quasi-non-interacting limit is reached at OD~$\lesssim 1$. 
It was not possible to work below this limit because of an increasing fraction of cells trapped in circular trajectories (due to hydrodynamic interactions with walls of the sample chambers, see \citep{LaugaCircle}) that did not explore the whole sample compartment, hindering relaxation towards a steady state. 
We therefore worked at OD~$\geq 1$. We report first data for OD = 5 ($\rho \approx 5\times 10^{9}$~cells/ml; $\phi \approx 0.7\%$) before discussing OD~=~1, where the data are noisier due to lower cell numbers.

%%%%%%%%%%%%%%%%%%%%%%%%%%%%%%%%%%%%%%%%%%%%%%%%%%%%%%%%%%%%%%%%%%%%
%\subsection{Qualitative test}

%%%%%%%%%%%%%%%%%%%%%%%%%%%%%%%%%%%%%%%%%%%%%%%%%%%%%%%%%%%%%%%%%%%%
\subsection{Stepped light pattern at OD = 5}

A field of AD10 cells rendered stationary by oxygen exhaustion was uniformly illuminated for $\approx\SI{20}{\minute}$ to achieve saturation speed  \citep{Jochen2018}. The light was then attenuated to $\mathcal{I}_-$, the level of the darker half of the target pattern, Eq.~\ref{eq:Ix}, for \SI{5}{\minute} to determine $\bar v_- = \SI{6.5(2)}{\micro\meter\per\second}$. Returning the intensity to its initial level, we waited another \SI{5}{\minute} for the swimming speed to return to $\bar v_+ = \SI{13.2 \pm 0.2}{\micro\meter\per\second}$. We measured the cell density $\rho_0$ and non-motile fraction $\beta_0$ of this high-speed uniform sample, and then switched on a stepped pattern, Eq.~\ref{eq:Ix}, by reducing the intensity in the $x > 0$ half plane. 

%......................................................
\begin{figure}[t]
\begin{center}
\includegraphics[width=0.55\columnwidth]{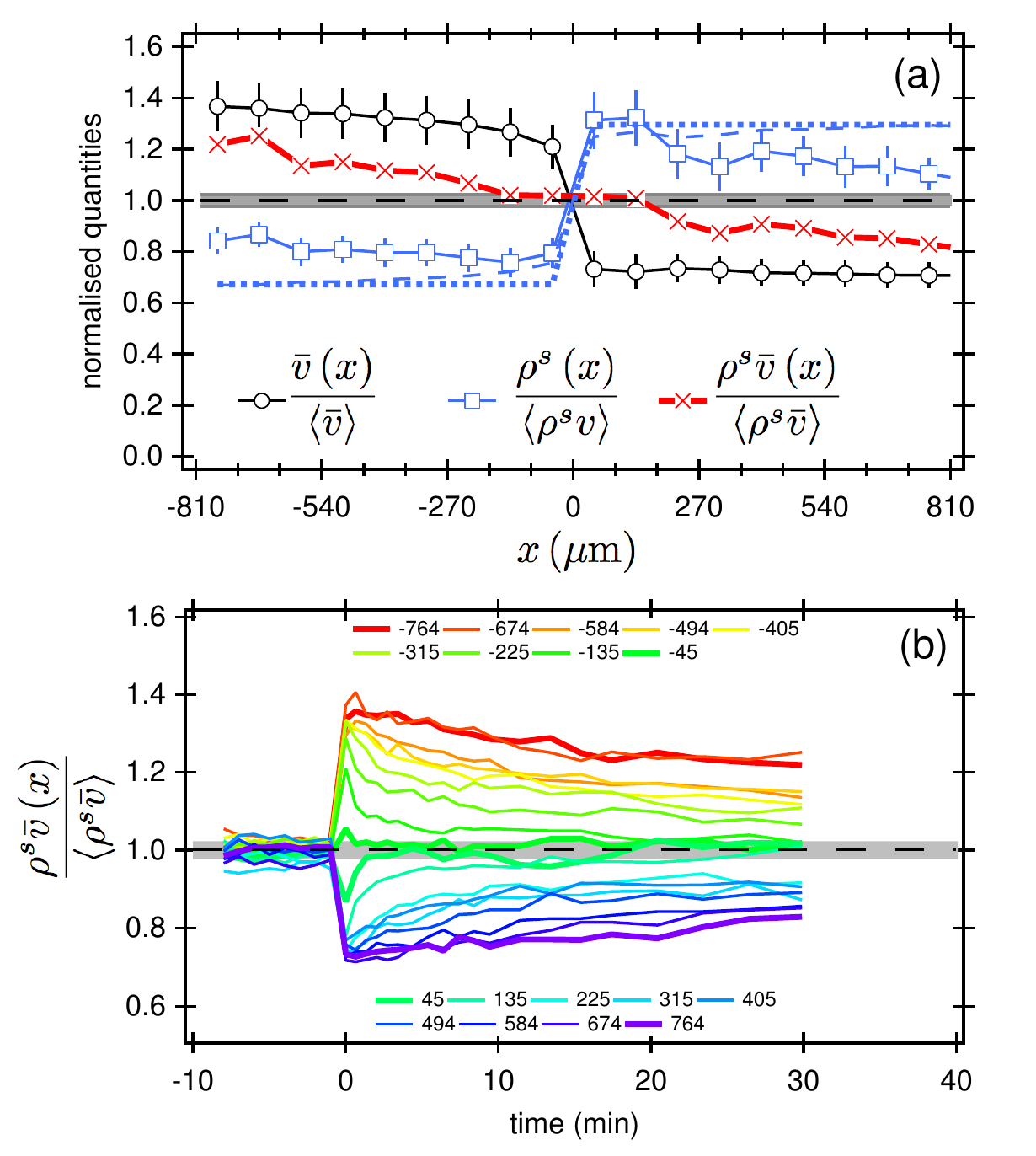}
\caption{{\label{fig:AD10OD5}
Response of AD10 at OD = 5.3 to a stepped intensity pattern: (a) normalised quantities as indicated {\it vs.}~$x$ after $\approx \SI{30}{\minute}$ of stepped illumination, and predictions for normalised $\rho^{\rm s}$ based on two speeds (dotted line) and complete measured $v(x)$ profile (dashed line). (b) Time dependence of normalised $\rho^{\rm s} v$ at various $x$ (see legend). The sample was uniformly illuminated before the halves pattern was switched on at $t = 0$. In both plots the grey area corresponds to the standard deviation in experimental values for uniform illumination ($t < 0$).
}}
\end{center}
% Data from 10/06/16 using AD10
\end{figure}
%......................................................

Figure~\ref{fig:AD10OD5}(a) shows the mean swimming speed averaged over $y$ tiles, $\langle \bar v(x,y) \rangle_y = \bar v(x)$, normalised to the whole-sample-averaged speed, $\langle v \rangle$, plotted against $x$ at \SI{30}{\minute} after switching on the stepped pattern. A stepped speed pattern developed, Fig.~\ref{fig:AD10OD5}a ($\circ$).

If Eq.~\ref{eq:rhoV} is valid, we expect the swimmer density to obey
\begin{equation}
\rho^{\rm s}_+(x) \bar v_+(x) = \rho^{\rm s}_-(x) \bar v_-(x), \label{eq:rhoSV}
\end{equation}
where `$\pm$' subscripts having their obvious meanings. If the density of non-motile cells is constant throughout the experiment (see SI for justification), i.e.
\begin{equation}
\rho_{\pm}^{\rm nm} = \beta_0\rho_0, \label{eq:rhoNM}
\end{equation}
we can write the total cell density on the two sides of $x= 0$ as
\begin{equation}
\rho_{\pm} = \rho_\pm^{\rm s}  + \rho_{\pm}^{\rm nm} = \rho_\pm^{\rm s}  + \beta_0 \rho_0.
\end{equation}
Finally, the average cell density is  
\begin{equation}
\rho_0 = \frac{1}{2}(\rho_+ + \rho_-). \label{eq:rho0}
\end{equation}
Equations~\ref{eq:rhoSV}-\ref{eq:rho0} together predict the density of motile cells in the two half planes: 
\begin{equation}
\rho_{\pm}^{\rm s} = \frac{2\rho_0(1-\beta_0)}{1 + \frac{v_{\pm}}{v_{\mp}}}. \label{eq:rhoPM}
\end{equation}

%......................................................................
\begin{figure}[t]
\begin{center}
\includegraphics[width=0.55\columnwidth]{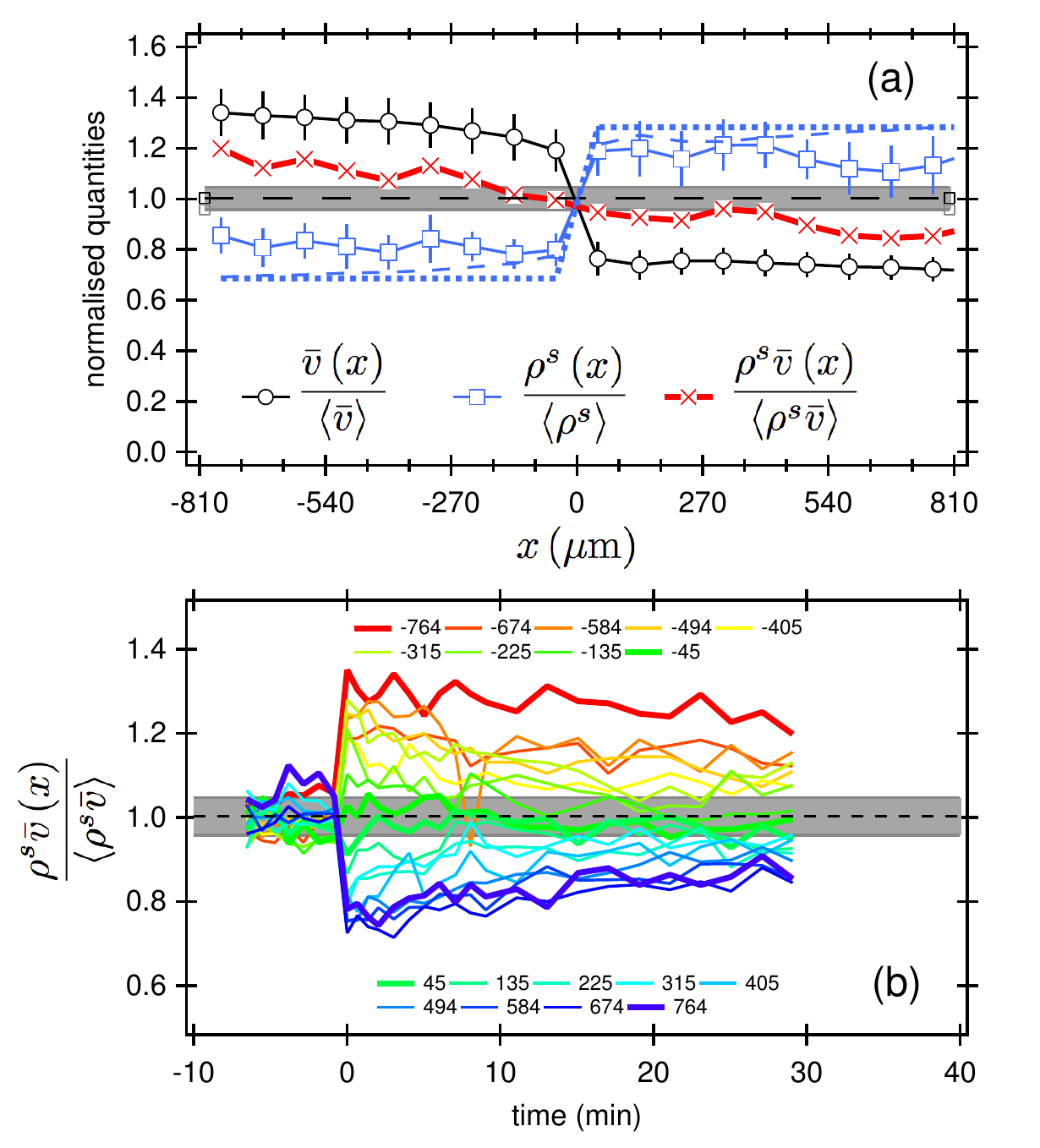}
\caption{Response of AD10 at OD = 1 to a stepped intensity pattern, showing the same quantities as Fig.~\ref{fig:AD10OD5}. Qualitatively the behaviour is the same as for the higher density, but the data are noisier due to the overall weaker signal. (See Figs.~\ref{fig:AD10OD1-maps} \& \ref{fig:AD10OD1-profiles} for spatial maps and time evolutions.)  {\label{fig:rhoSV}
}}
\end{center}
% Data from 
\end{figure}
%......................................................................

We calculated the swimmer density in our experiments from the measured total cell density $\rho(x)$ and non-motile fraction $\beta(x)$ using $\rho^{\rm s}(x) = \rho(x)[1-  \beta(x)]$, and normalised it by the whole-sample-averaged swimmer density. This data \SI{30}{\minute} after the imposition of the stepped intensity pattern is also stepped, Fig.~\ref{fig:AD10OD5}a ({\color{blue} $\square$}), with the theoretical predictions from Eq.~\ref{eq:rhoPM} using the measured average $v_\pm$ as inputs, Fig.~\ref{fig:AD10OD5}a (dotted line), giving a reasonable account of the step amplitude. A more sophisticated version of this model which takes the measured shape of $\bar v(x)$ into account (see SI for details) is able to capture the amplitude of the jump in $\rho^{\rm s}(x)$ at $x = 0$ even more precisely, Fig.~\ref{fig:AD10OD5}a (dashed line).

The product $\rho^{\rm s}\bar v(x)$ normalised to the whole sample average, Fig.~\ref{fig:AD10OD5}a ({\color{red} $\times$}), is indeed constant for $|x| \lesssim \SI{200}{\micro\meter}$, verifying Eq.~\ref{eq:rhoSV}, which is the application of Eq.~\ref{eq:rhoV} to our conditions. 
However, there are systematic deviations from constancy at $|x| \gtrsim \SI{200}{\micro\meter}$.  
One possible explanation is the emergence of collective motion with associated local nematic ordering \cite{Kessler2011}, which would invalidate the derivation of Eq.~\ref{eq:rhoV}. However, we only observed collective motion at OD~$\gtrsim 10$. 
Instead, the deviations of $\rho^{\rm s} \bar v$ from constancy at $|x| \gtrsim \SI{200}{\micro\meter}$ is a kinetic effect. Figure~\ref{fig:AD10OD5}b shows the time evolution of the normalised $\rho^{\rm s} \bar v(x)$ at different~$x$. Steady state was reached rapidly for $|x| \lesssim \SI{200}{\micro\meter}$, but was not reached by \SI{30}{\minute} at the extremes of our observation window, $|x| \gtrsim  \SI{600}{\micro\meter}$. Given their effective diffusivity $D_{\rm eff} \sim 10^3\,\si{\square\micro\meter\per\second}$, cells at the extremities of our compartment take $\gg \SI{30}{\minute}$ to sufficiently sample both speed regions, preventing the attainment of steady state within our observational time window. 
This leads to the deviations between observed and predicted $\rho^{\rm s}(x)$ away from $x = 0$.
Nevertheless, Figure~\ref{fig:AD10OD5}b suggests that $\rho^{\rm s}\bar v =$~constant should be attained at all $x$ in the long-time limit.

%%%%%%%%%%%%%%%%%%%%%%%%%%%%%%%%%%%%%%%%%%%%%%%%%%%%%%%%%%%%%%%%%%%%
\subsection{Stepped pattern at other cell densities} 

\begin{figure}[t]
\begin{center}
\includegraphics[width=0.5\columnwidth]{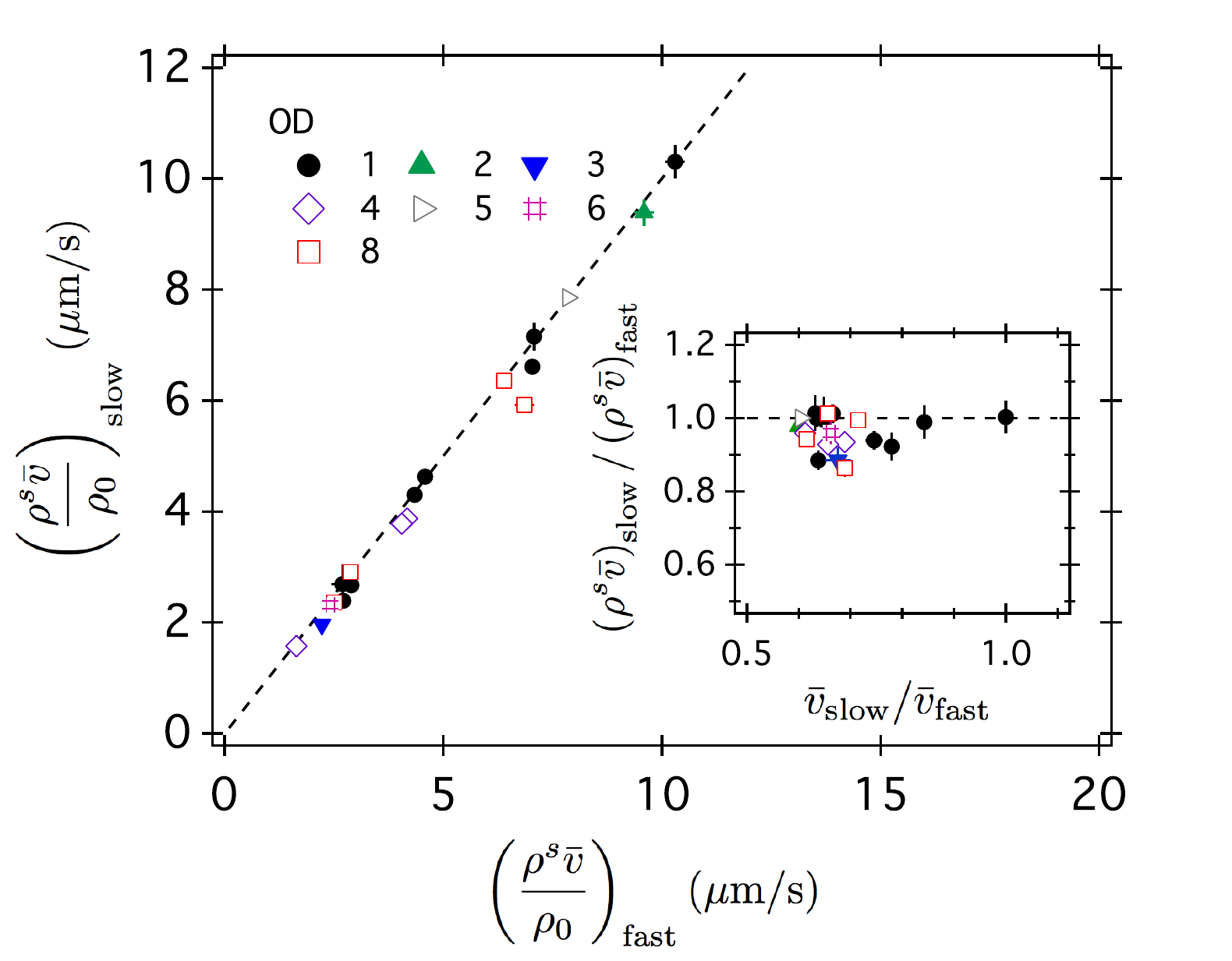}
\caption{A plot of $\rho^{\rm s} \bar v$ for the first tile on the fast side vs.~the same quantity for the first tile on the slow side for several independent data sets. For both low density (filled symbols) as well as higher density data sets (open symbols) eq.~[\ref{eq:rhoSV}] holds. This is also demonstrated in the inset, which shows $(\rho^{\rm s}_- \bar{v}_-/(\rho^{\rm s}_+ \bar{v}_+)$ vs. $\bar{v}_-/\bar{v}_+$ for the same datasets. {\label{fig:highdensity}
}}
\end{center}
% Data from 
\end{figure}

Measurements and model predictions for the lower OD = 1 are shown in Fig.~\ref{fig:rhoSV}(a, b). The data are noisier, but show the same trends. In the vicinity of $x = 0$, $\rho^{\rm s}\bar v \approx$~constant. To highlight the behaviour in the two \SI{90}{\micro\meter}-wide stripes of tiles bordering $x = 0$, we plot  $\rho^{\rm s}\bar v(x)$ at $t = \SI{30}{\minute}$ for these two stripes against each other for a number of independent experiments, Fig.~\ref{fig:highdensity} ($\bullet$). In all cases, $\rho^{\rm s}\bar v(x) =$~constant for these central stripes to within experimental uncertainties. The ratio of $\rho^{\rm s}\bar v(x)$ in these two stripes plotted against the ratio of the swimming speed on the two half planes, Fig.~\ref{fig:highdensity} inset ($\bullet$), is consistent with this claim.

We performed experiments using the stepped light pattern at other cell densities and also using an additional smooth swimming strain (DM1). In all cases up to OD = 8, we find that $\rho^{\rm s}\bar v$ is constant across the two central stripe of tiles on either side of $x = 0$, Fig.~\ref{fig:highdensity}, where we are certain that steady state has been reached, verifying Eq.~\ref{eq:rhoV} up $\rho \approx 8 \times 10^9$~cells/ml.

%%%%%%%%%%%%%%%%%%%%%%%%%%%%%%%%%%%%%%%%%%%%%%%%%%%%%%%%%%%%%%%%%%%%
\subsection{Measurements using a periodic light pattern}

The  complicating factor so far is the slow global convergence towards $\rho^{\rm s}(x) v(x) = \textrm{const}$, so that steady state will only be reached in $\sim$~hours. Experiments on such time scales are impractical due to mechanical and biological stability issues. Thus, we only have direct evidence for the validity of Eq.~\ref{eq:rhoV} in the vicinity of the intensity step at $x = 0$. This suggests that the use of a series of thin stripes would give more clear-cut results unencumbered by kinetic issues. We found that this was indeed the case. 

\begin{figure}[t]
\begin{center}
\includegraphics[width=0.55\columnwidth]{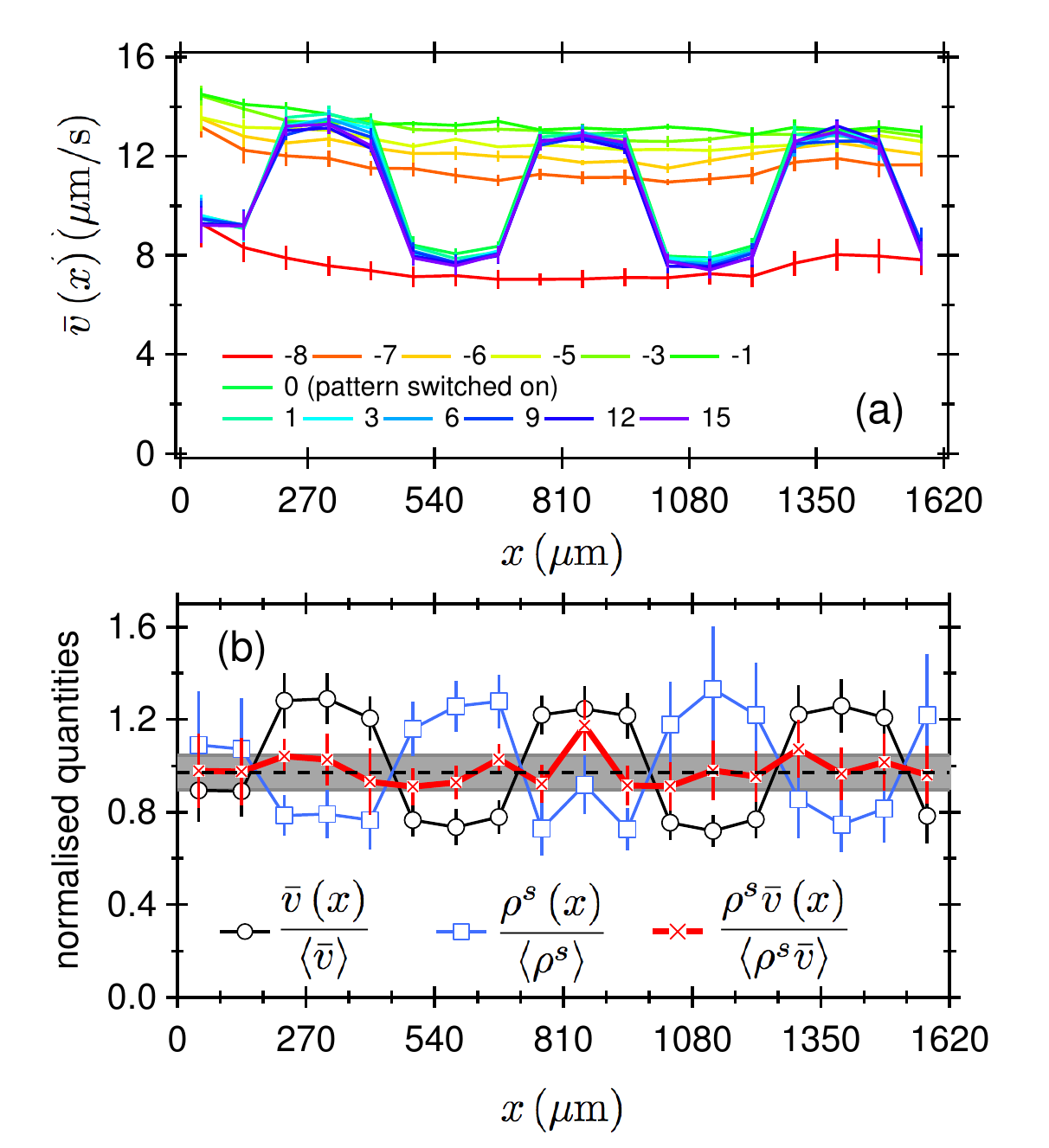}
\caption{{\label{fig:lines}
Response of bacteria to the imposition of a 1D square wave with $\SI{270}{\micro\meter}$ wide stripes. (a) The evolution of the speed profile normalised to its (time) average, with time increasing from red to violet (legend gives $t$ in minutes): initially illumination is uniform, with the intensity getting increased from low to high at $t=-\SI{7}{\minute}$, then the pattern is applied for \SI{15}{\minute}. (b) The spatial profiles of various normalised quantities as indicated in the legend after \SI{15}{\minute} of patterned illumination. Error bars represent SD. Dashed line (and grey area) is the average ($\pm$ 1 SD) over all tiles for the uniform case ($t<0$). 
}}
\end{center}
% Data from 23/09/16
% AD10 sample, OD1
% Analysis in Analysis\2016\Q3\160923-AD10\160923-AD10-OD1_Summary_All_5rows.pxp
% Plotted results are for 00_009
% Last uniform sample 00_005
\end{figure}

In response to the imposition of a one-dimensional square-wave illumination pattern of brighter-darker stripes with \SI{540}{\micro\meter} repeat generated by a digital mirror device \citep{Jochen2018}, the swimming speed of bacteria changed from a uniform distribution to a square-wave distribution almost instantaneously (in $\lesssim \SI{1}{\second}$), Fig.~\ref{fig:lines}(a). 
This in turn modified the cell density (initially uniform at OD = 1), which approached a steady state much more quickly. This is possible not only because of the length scale reduction, but also because swimmers can enter (say) a high-intensity region from low-intensity regions on both the left and right.

Fig.~\ref{fig:lines}b shows the normalised swimmer density $\rho^{\rm s}(x)/\langle \rho^{\rm s} \rangle$ after \SI{6}{\minute} of patterned illumination, together with $\bar v(x)/\langle v \rangle$ and their product. While the data are  again somewhat noisy because of the low average cell density (OD = 1), it is clear that $\rho^{\rm s}\bar v=$~constant to within one standard deviation, which directly verifies Eq.~\ref{eq:rhoV}. 

%......................................................................
%......................................................................

%%%%%%%%%%%%%%%%%%%%%%%%%%%%%%%%%%%%%%%%%%%%%%%%%%%%%%%%%%%%%%%%%%%%
\subsection{Experiments with $\beta(I)$ dependency}

Interestingly, experiments using low light intensities (which gave low swimming speeds) proved less successful, because at very low intensities we found a noticeably higher percentage of non--motile cells in the sample than at high light intensities (Fig.~\ref{fig:betaInt}).
This led to a spatial variation in the non-motile density (Fig.~\ref{fig:AD4star_1605_12_profiles}), which considerably complicates the interpretation and analysis of such experiments (see SI for details).

%%%%%%%%%%%%%%%%%%%%%%%%%%%%%%%%%%%%%%%%%%%%%%%%%%%%%%%%%%%%%%%%%%%%
\subsection{Experiments using light-powered run-and-tumble strain}

We end by explaining why we did not use motility wild type (run and tumble) strains for our experiments. Their motion randomises much more rapidly than smooth swimming mutants, which would have significantly alleviated the non-steady-state issue for the stepped intensity pattern. However, we found that AD4, a PR-bearing motility wild-type, gathered near $x = 0$, 
%where $d\mathcal{I}(x)/dx$ is highest 
on the darker side of the intensity step (see Figs.~\ref{fig:AD4-maps} \& \ref{fig:AD4-profiles}). From the $q$ dependence of our DDM data \citep{MartinezDDM} we can deduce that the tumbling rate increases noticeably as cells swim from light to dark, whereas cells swimming from dark to light do not show any obvious change in their tumbling behavior. This may be due to `energy taxis' \citep{Schweinitzer2010}. The validity of Eq.~\ref{eq:rhoV} depends on the assumption that the tumbling rate is independent of swimming direction \citep{Schnitzer1993,TailleurPRL}, so that motility wild types cannot be used to test this result.

%%%%%%%%%%%%%%%%%%%%%%%%%%%%%%%%%%%%%%%%%%%%%%%%%%%%%%%%%%%%%%%%%%%%
\section*{Summary and Conclusions}

Equation~\ref{eq:rhoV} is one of only a handful of exact predictions to date on the statistical mechanics of active particle systems. Its `weak' form, for non-interacting systems was derived for RTPs \cite{Schnitzer1993}, while its `strong' form was later derived both for RTPs \cite{TailleurPRL} and ABPs \cite{Cates2016}. Taken together, our experiments using stepped and stripped light patterns give strong evidence that Eq.~\ref{eq:rhoV} holds at $1 \leq$~OD~$\leq 8$ ($0.15\% \lesssim \phi \lesssim 1.2\%$) for smooth-swimming {\it E. coli} whenever we can be sure that steady state has been reached, either in the vicinity of $x = 0$ in the stepped pattern or throughout the stripped pattern. Our swimmers are interacting throughout our concentration range \cite{Morozov2017}, even though collective motion is not observed until OD~=~10. Thus, our results verify the `strong' form of Eq.~\ref{eq:rhoV} for ABPs.

The qualitative validity of Eq.~\ref{eq:rhoV}, viz., that cells gather where they swim slower, or, equivalently for our cells, where the light intensity is lower, has already been assumed and utilised in recent work deploying such cells in smart (or reconfigurable) templated self assembly, or `painting with bacteria' \citep{Jochen2018,Frangipane2018}. Indeed, in a recent demonstration of how to perform bacterial painting with multiple shades of graded intensity levels \citep{Frangipane2018}, there was attempt at checking the correctness of Eq.~\ref{eq:rhoV} {\it en passant}, which, however, was unsuccessful because of a high number of non-motile cells and the long stopping time of their strain, the latter producing memory effects. Our success in verifying Eq.~\ref{eq:rhoV} shows that carefully quantifying and subtracting the non-motile fraction and the use of a strain of bacteria with very short stopping time are essential ingredients in such an experiment. 
Indeed, without careful design most `real' active systems are likely to display dynamic behaviour which is too complex to fulfill the assumptions leading to eq.~\ref{eq:rhoV}, as evidenced by our findings for the run-and-tumble strain.
However, our experimental method of spatially resolved DDM can reliably quantify swimming speed and relative density (along with many other parameters) even in such cases. As such it can provide new insights into a wide variety of sytems displaying spatially varying dynamics, from biological taxis studies \citep{Rosko2017} to collective motion.

Throughout, we have focussed on steady-state effects, although the consideration of time dependence proved crucial in interpreting apparent systematic deviations from the prediction of Eq.~\ref{eq:rhoV} for imposed stepped intensity patterns. Time-dependent effects are, of course, interesting in their own right. Thus, the response of active particles to a time-dependent topographic landscape that is self-assembled by the cells themselves \citep{Cates2016} has yet to be explored experimentally. On the other hand, it has recently been suggested theoretically \citep{Maggi2018} and demonstrated experimentally \citep{Koumakis2018} that travelling-wave light fields can be exploited for transporting and rectifying light-activated swimmers. Exploitation of these and other opportunities should open up new fields of fundamental studies and applications.

%%%%%%%%%%%%%%%%%%%%%%%%%%%%%%%%%%%%%%%%%%%%%%%%%%%%%%%%%%%%%%%%%%%%
\section*{Materials and Methods}

\label{sec:material}
\vspace{-0.5cm}
\subsection*{Cells}
We constructed 3 different strains of {\it E.~coli} using plasmids expressing SAR86 $\gamma$-proteorhodopsin (a gift from Jan Liphardt, UC Berkley).
These strains were designed to exhibit a fast response to changes in light intensity. This was achieved by deleting the {\it unc} gene cluster, so that the F$_1$F$_0$-ATPase membrane protein complex cannot work in reverse in the dark to generate proton motive force to power swimming. The detailed molecular biology and strain characterisation have been reported before \citep{Jochen2018}. AD4 is a WT (run-and-tumble) swimmer derived from AB1157, whereas DM1 and AD10 are smooth swimming strains derived from RP437 and AB1157, respectively (see SI table~\ref{tab:strains}).
%AD4 was found unsuitable because its tumble rate is highly intensity dependent. 
The 2 smooth swimming strains behaved similarly, although AD10 achieved a much higher swimming speed than DM1 and was also more efficiently powered by light.
Therefore we mostly used AD10, with some additional data acquired using DM1.

Overnight cultures were grown aerobically in 10 mL Luria-Bertani Broth (LB) using an orbital shaker at 30$^\circ$C and 200~rpm. A fresh culture was inoculated as 1:100 dilution of overnight grown cells in 35ml tryptone broth (TB) and grown for $\SI{4}{\hour}$ to an optical density of OD$_{600}\approx0.2$. The production of proteorhodopsin (PR) was induced by adding arabinose to a concentration of 1~mM as well as the necessary cofactor all-trans-retinal to $10~\mu$M to the growth medium.
Cells were incubated under the same conditions for a further hour to allow protein expression to take place and then transferred to motility buffer (MB, pH = 7.0, 6.2 mM K$_2$HPO$_4$, 3.8 mM KH$_2$PO$_4$, 67 mM NaCl and 0.1 mM EDTA).
Single filtration (0.45~$\mu$m HATF filter, Millipore) was used to prepare high density stock solutions (OD$\approx 8$) which were diluted with MB to the desired cell concentration.

The samples were loaded into commercial \SI{2}{\micro\liter} sample chambers (SC-20-01-08-B, Leja, NL) of dimensions $\approx \SI{6}{\milli\meter} \times \SI{10}{\milli\meter} \times \SI{20}{\micro\meter}$, where cells predominantly swim in the $(x,y)$ (imaging) plane, but have enough room to `overtake' each other in all three spatial dimensions. The chamber was then sealed using vaseline to stop air flow, so that swimming stopped once dissolved oxygen was exhausted \citep{Poon2016a}. This happened in $\approx 20$~min at OD$~\approx 1$ ($\approx 10^9$ cells/ml or $0.2\%$ volume fraction of cell bodies). Thereafter, we controlled the activity of the cells  by illuminating with green light of various intensities \citep{Jochen2018}. 

%%%%%%%%%%%%%%%%%%%%%%%%%%%%%%%%%%%%%%%%%%%%%%%%%%%%%%%%%%%%%%%%%%%%
\subsection*{Experimental setup}
\label{sec:microscope}
The samples were observed using a Nikon TE2000 inverted microscope with a PF~$10 \times$, N.A. 0.3 phase contrast objective. Time series of movies ($\sim \SI{40}{\second}$ duration at 100 frames per second) were recorded using a CMOS camera (MC~1362, Mikrotron). A long-pass filter (RG630, Schott Glass) in the bright-field light path ensured that the imaging light did not activate PR. The light controlling bacterial swimming was provided by an LED (Sola SE II, Lumencor) whose intensity was set via a computer interface. The LED light was filtered to a green wavelength range ($510$ -- $\SI{560}{\nano\meter}$) overlapping with the absorption peak of our PR \citep{Walter2007} and illuminated the sample in a trans-illumination geometry. By illuminating an area $\gg$ the field of view of the objective,  we minimised the loss of swimmers over time. If only a small region of the sample is illuminated, the density of swimmers continuously drops, because they reach the illumination boundaries and accumulate there (no light = no swimming). For the stepped pattern experiment we uniformly illuminated a $\approx \SI{7}{\milli\meter}$ diameter circle, covering almost all of the sample chamber. Under these conditions the cell density is conserved, thus simplifying theoretical modeling. We used a thin sheet polariser imaged onto the sample plane to attenuate the intensity on half of the sample. A digital mirror device~\citep{Jochen2018} projected the periodic pattern onto a $\approx \SI{2.9}{\milli\meter}$ diameter area of the sample. 

%%%%%%%%%%%%%%%%%%%%%%%%%%%%%%%%%%%%%%%%%%%%%%%%%%%%%%%%%%%%%%%%%%%%
\subsection*{Differential Dynamic Microscopy}

DDM measures $(\bar v,  \beta)$ averaged over $10^4 - 10^5$ cells under our conditions \citep{WilsonDDM, MartinezDDM, Poon2016a}. From $\approx \SI{30}{\second}$ of wide-field, low-magnification movies, one extracts the power spectrum of the difference between pairs of images delayed by time $\tau$, $g(\vec{q},\tau)$, where $\vec{q}$ is the spatial frequency vector. Under suitable conditions and for isotropic motion, the intermediate scattering function $f(q, \tau)$, the $q^{\rm th}$ mode of the density autocorrelation function, is given by:
\begin{equation}
  g(q,\tau) = A(q)\left[1 - f(q,\tau)\right] + B(q) \,. \label{eq:gQtau}
\end{equation} 
Here, $B(q)$ relates to the background noise and $A(q)$ is the signal amplitude. Fitting $f(q,\tau)$ to a suitable swimming model of {\it E. coli} yields 4 key motility parameters: the mean $\bar v(q)$, and width $\sigma(q)$ of the speed distribution $P(v)$, the non-motile fraction $\beta(q)$, and the diffusion coefficient of non-motile cells $D(q)$, as a function of $q$. All of these should, ideally, be $q$-independent. In practice, there is some $q$-variation. We typically averaged the fitting parameters over $0.5<q<2.2\,\mu\text{m}^{-1}$ to give, e.g. $\bar v=\left<v(q)\right>_q$ and $\beta=\left<\beta(q)\right>_q$.

In a dilute system whose structure factor $S(q) \approx 1$, $A(q)$ is proportional to the sample density \citep{Reufer,ConfocalDDM} and can therefore be used to determine relative densities by $\rho_1/\rho_0 = \left< A_1(q)/A_0(q) \right>_q$  \citep{Jochen2018}. Note that ratioing the $A(q)$s removes their strong $q$-dependence. 

%%%%%%%%%%%%%%%%%%%%%%%%%%%%%%%%%%%%%%%%%%%%%%%%%%%%%%%%%%%%%%%%%%%%
%\subsection*{Spatially-resolved DDM}

Spatially-resolved DDM is in principle straightforward: the above algorithm simply needs to be implemented on $p\times p$ (pixel)$^2$ sub-movies. In practice, care is required in choosing the minimum $p$ for which meaningful results can be obtained. We do this by illuminating a field of cells uniformly, measuring $(\bar v, \beta, \rho, \rho^{\rm s} v)$ from individual $p \times p$ tiles in the steady state, and obtaining the probability distribution of these parameters. Under our imaging conditions, we found that these distributions became $p$-independent when $p \geq 64$. We therefore chose $p = 64$, corresponding to \SI{90}{\micro\meter} in the sample (see SI \S\ref{SI:sDDM} for details). 

A full $512\times512$ movie yields $g(q,\tau)$s at $512/2= 256$ distinct $q$ values. We divide it into 64 sub-movies of size $64\times64$ (pixel)$^2$. This yields $8 \times 8 \times (64/2) = 2048$ $g(q,\tau)$s  to be fitted to give for each sub-movie $v_{x,y}(q)$, $\beta_{x,y}(q)$, and $\rho_{x,y}(q)/\rho_0=A(q)_{x,y}/A_0(q)$, where $A_0(q)$ is measured from the same sample under uniform illumination (i.e. just before switching to a structured light pattern) averaged over $0.5\leq q \leq 1.5\,\mu\text{m}^{-1}$. 
The upper $q$ limit is somewhat lower than what typical for whole-movie analysis \citep{MartinezDDM} due to non-systematic failure of fitting at higher $q$ values, presumably due to noise or windowing artefacts \citep{Giavazzi2017}.
% (See SI \S\ref{SI:sDDM} for details.)

\subsection*{Data availability} The data presented here is available on the Edinburgh DataShare repository \cite{Arlt2018:rhoV:DS}.

%%%%%%%%%%%%%%%%%%%%%%%%%%%%%%%%%%%%%%%%%%%%%%%%%%%%%%%%%%%%%%%%%%%%

% Bibliography
\bibliography{active}

\subsection*{Acknowledgements}
We were funded by the EPSRC (EP/J007404/1) and the ERC (AdG 340877 PHYSAPS). We thank Mike Cates, Julien Tailleur, Alexander Morozov and Nick Koumakis for helpful discussions, Jan Liphardt for a gift of PR plasmids and Dario Miroli for {\it E. coli} strain DM1.

\subsection*{Author Contributions}
J.A. and V.A.M. contributed equally to this work. WCKP initiated the work. TP and AD designed mutants constructed by AD. JA and VAM performed experiments, analysed and interpreted data with TP and WCKP, and wrote manuscript with WCKP.

\subsection*{Competing Interests} The authors declare that they have no competing financial interests.

\subsection*{Correspondence} Correspondence and requests for materials should be addressed to J.~Arlt~(email: j.arlt@ed.ac.uk).
%******************
\end{document}